\documentclass[preprint]{arxiv}

\usepackage{amsthm,amsmath}
\RequirePackage[colorlinks,citecolor=blue,urlcolor=blue]{hyperref}
\usepackage{epsfig, verbatim}
\usepackage{graphics, epstopdf}
\usepackage{multirow}
\usepackage{array}
\usepackage{soul}
\usepackage{float}

\startlocaldefs
\newcolumntype{L}[1]{>{\raggedright\let\newline\\\arraybackslash\hspace{0pt}}m{#1}}
\newcolumntype{C}[1]{>{\centering\let\newline\\\arraybackslash\hspace{0pt}}m{#1}}
\newcolumntype{R}[1]{>{\raggedleft\let\newline\\\arraybackslash\hspace{0pt}}m{#1}}

\newcommand{\bc}{ { \bf c }}
\newcommand{\bD}{ { \bf D }}

\newcommand{\bI}{ { \bf I }}
\newcommand{\bm}{ { \bf m }}
\newcommand{\bR}{ { \bf R }}

\newcommand{\bt}{ { \bf t }}
\newcommand{\bV}{ { \bf V }}
\newcommand{\by}{ { \bf y }}
\newcommand{\bY}{ { \bf Y }}
\newcommand{\bepsilon}{\mbox{\boldmath{$\epsilon$}}}

\newcommand{\bSigma}{\mbox{\boldmath{$\Sigma$}}}

\newcommand{\btheta}{\mbox{\boldmath{$\theta$}}}
\newcommand{\bzero}{\mbox{\boldmath{$0$}}}
\newcommand{\bone}{\mbox{\boldmath{$1$}}}

\newcommand{\pr}{ \mathrm{Pr} }

\endlocaldefs

\begin{document}

\begin{frontmatter}

\title{A Bayesian Nonparametric Model for Predicting Pregnancy Outcomes Using Longitudinal Profiles}
\runtitle{BNP Classification of Longitudinal Data}

 \author{\fnms{Jeremy T.} \snm{Gaskins}\corref{}\ead[label=e1]{jeremy.gaskins@louisville.edu}\thanksref{m1},
\fnms{Claudio} \snm{Fuentes}\ead[label=e2]{fuentesc@stat.oregonstate.edu}\thanksref{m2}, and
\fnms{Rolando} \snm{De La Cruz}\ead[label=e3]{rolando.delacruz@pucv.cl}\thanksref{m3}
 }
\affiliation{University of Louisville\thanksmark{m1}, Oregon State University\thanksmark{m2}, and Pontificia Universidad Cat\'olica de Valpara\'{i}so\thanksmark{m3}}

\address{\thanksmark{m1}University of Louisville\\
School of Public Health and Information Sciences\\
Department of Biostatistics and Bioinformatics\\
Louisville, KY 40202, USA\\
\printead{e1}}

\address{\thanksmark{m2}Oregon State University\\
Department of Statistics\\
Corvallis, OR 97331, USA\\
\printead{e2}}

\address{\thanksmark{m3} Pontificia Universidad Cat\'olica de Valpara\'{i}so\\
Instituto de Estad\'{i}stica\\
Valpara\'{i}so, Chile\\
\printead{e3}}


\runauthor{J. Gaskins et al.}

\begin{abstract}
Across several medical fields, developing an approach for disease classification is an important challenge.  The usual procedure is to fit a model for the longitudinal response in the healthy population, a different model for the longitudinal response in disease population, and then apply the Bayes' theorem to obtain disease probabilities given the responses.  Unfortunately, when substantial heterogeneity exists within each population, this type of Bayes classification may perform poorly.  In this paper, we develop a new approach by fitting a Bayesian nonparametric model for the joint outcome of disease status and longitudinal response, and then use the clustering induced by the Dirichlet process in our model to increase the flexibility of the method, allowing for multiple subpopulations of healthy, diseased, and possibly mixed membership.  In addition, we introduce an MCMC sampling scheme that facilitates the assessment of the inference and prediction capabilities of our model.  Finally, we demonstrate the method by predicting pregnancy outcomes using longitudinal profiles on the $\beta$--HCG hormone levels in a sample of Chilean women being treated with assisted reproductive therapy.
\end{abstract}



\end{frontmatter}


\section{Introduction}\label{sec:intro}
The human chorionic gonadotropin beta subunit $\beta$--HCG increases in concentration during the early stages of pregnancy and is commonly used in obstetrics as a measure of pregnancy evolution. In particular, in the context of 
assisted reproductive therapy,
it is used as a prognostic marker to detect ectopic pregnancies and other possible complications \cite{obs1, obs2}. 
In a normal pregnancy, the level of $\beta$--HCG will roughly double every 1.5 days up to 5 weeks following fertilization, and then every 3.5 days starting on the 7th week \cite{obs3}. 
After  the first trimester, levels gradually decrease during the remainder of the pregnancy and quickly drop to zero once the pregnancy has ended. 
On the other hand, ectopic pregnancies, miscarriages or spontaneous abortions often exhibit a decrease or a low rate increase in the $\beta$--HCG levels during the first trimester. However, other complications can be preceded by an abrupt rise of hormone concentration, 
and consequently, any efforts to predict an adverse pregnancy condition based on the $\beta$--HCG level needs to take into account the dynamics of the concentration over time and not only the absolute values at a given moment. 

In this paper, we focus on the detection of possible pathologies during pregnancy, using the longitudinal profiles of the hormone concentrations in pregnant women when no other covariate information is available. 
Since  hormone measurements are recorded infrequently and at different stages of pregnancy for every woman, actual data sets often consist of sparse, unbalanced, and highly irregular longitudinal observations (see Figure {\color{blue}1}). 
In this context, some recent approaches, in which individuals in all groups are modeled jointly and estimation and classification are performed simultaneously, have been discussed in the literature using functional data analysis (FDA) \cite{Yao05,Yao05b}. 
For instance, \cite{muller:05} considered a functional binary regression model for sparse functional data, and \cite{lutsetal:12} proposed a least squares support vector machine classifier for longitudinal data. More recently, \cite{YaoWuZou:16} considered the projection of the sparse functional data onto the most effective directions associated with a functional index model using a weighted support vector machine and proposed a cumulative slicing approach to borrow information across individuals.

The analysis of the pregnancy data set we introduce in Section \ref{data} has motivated the introduction of a number of classification methods for sparse longitudinal data using mixed models. Due to their flexibility, the use of such models, in which individual trajectories can borrow information from each other, have proven useful in describing individual and group behaviors. For instance, \cite{MarshallBaronSIM2000} used a nonlinear mixed effects (NLME) model approach  to describe the  evolution in the different groups and produce an optimal allocation rule. The authors showed the necessity of modeling the interaction of time with fixed and random effects in a nonlinear way in order to capture the dynamics of the data set. In this direction, \cite{DelaCruzQuintana07} proposed a Bayesian framework for the classification of longitudinal profiles, when the underlying structure in each group or populations can be expressed by nonlinear hierarchical models. In \cite{DelaCruzQuintanaMuller07}, the authors extended these ideas and developed a semi-parametric Bayesian approach, in which the distribution of the random effects was estimated using a Dirichlet process mixture prior.

More recently, \cite{arribas2015classification} proposed a semi-parametric linear mixed--effects model (SLMM) for the longitudinal trajectory of each group and considered a LASSO approach to estimate the function capturing the temporal trend of the data for the semi-parametric component of the model. 
Although the use of a LASSO--type estimator was motivated by the irregularities observed in the abnormal group, 
 results suggested that the use of penalized splines, could be appropriate. 
Such an approach, incorporating low rank penalized splines into mixed-effects models is further explored in \cite{de2017predicting}, and an extension discussing a Bayesian estimation approach to the problem is introduced in \cite{de2016bayesian}. A study looking at the misclassification rates of some of these methods can be found in  \cite{de2016error}.


In this paper, motivated by the flexibility of  mixed--effect models, we consider the problem from a Bayesian perspective and introduce a Bayesian non-parametric (BNP) model for classification. These methods provide a flexible collection of alternatives that are useful in a variety of contexts, in particular when a clustering is desired (see \cite{gershman2012, hjort2010bayesian}, for an introduction to the topic).
In Section \ref{data} we introduce the data set that motivated this work, and in Section \ref{model} we specify the model and the hierarchical structure. In Section \ref{estimation} we introduce an MCMC sampling scheme and briefly discuss the inferential problem. In Section \ref{simulations} we present a numerical study to assess the main properties of our new approach, and in Section \ref{application} we implement our model to analyze the pregnancy data set. We finish with a few remarks in Section \ref{disc}.

\begin{figure}[t]
\includegraphics[width= 1\textwidth]{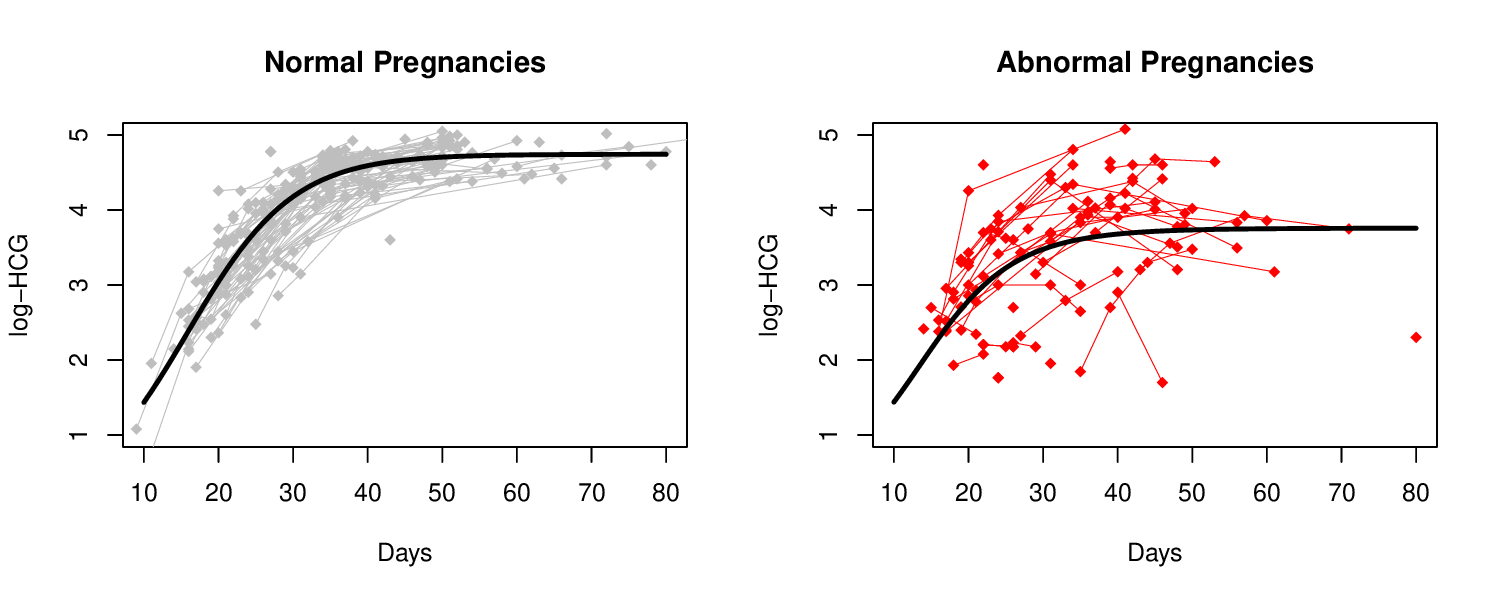}
\label{spaghetti}
\caption{Spaghetti plots of longitudinal trajectory by pregnancy outcome.  The solid line represents the best fitting trajectory in the 2-component model.}
\end{figure}

\section{Assisted Reproductive Therapy in Chilean Women}\label{data}
The data used in this paper were collected during a clinical trial in a private assisted reproduction center in Santiago, Chile. The data set consists of repeated measures of $\beta$--HCG concentration levels taken over a period of two years on 173 pregnant women.  Forty-nine pregnancies presented serious anomalies that ended up with the lost of the fetus; in the other 124, the pregnancy completed with a normal development that went to term without important complications.
The $\beta$--HCG concentration levels were recorded at different times for each women during the first trimester (days 10 to 80) of pregnancy.  
 Figure \ref{spaghetti}
 shows a clear differences in the trajectories of the $\beta$--HCG hormone (in log-scale) for women with \emph{normal} (successful) and \emph{abnormal} (unsuccessful) pregnancies.  

The data set is unbalanced, typically including only a few observations per subject, measured at irregular intervals over time . The number of measurements per woman ranges between one and six with a median of two. In the group of normal pregnancies, $28\%$ of the women  have only one measurement, and almost $98\%$ of the women have three or fewer measurements. For the abnormal group the proportions are  $35\%$ and $86\%$, respectively. The time between consecutive measurements (on the same subject) ranges from 2  to 51 days, with a large variability between subjects.

The profiles, in particular for the abnormal group, show an erratic behavior that is difficult to capture using a single model, even when applying adaptive techniques such as LASSO.
Unlike other studies of similar nature, no covariates are available for this data.

\section{Bayesian Nonparametric Model for Classification}\label{model}

For a study population of $N$ patients, let $D_i$ be a binary indicator of the disease status for patient $i=1,\ldots,N$, with $D_i=1$ for diseased patients, and $D_i=0$ for healthy (not diseased) patients.  The patient is observed on $n_i$ occasions at measurement times $\bt_i = (t_{i1},\ldots,t_{in_i})$.   We assume the measurement times are recorded with respect to some comparable start time. For instance, in the pregnancy application, $t=0$ represents the time of fertilization.  The longitudinal measurement times are not necessarily balanced; patients are followed for a varying number of times and at different time points. The $j$th measurement $Y_{ij}$ is recorded at time $t_{ij}$, and we let $\bY_i = (Y_{i1},\ldots,Y_{in_i})$.

For $i=1,\ldots,N$, we model the patient data $D_i$ and $\bY_i$, conditionally on a patient-specific parameter vector $(\phi_i,\btheta_i)$, according to the following structure
{\small
\begin{eqnarray}\label{m1}
D_i\mid\phi_i &\sim& \mbox{Bern}(\phi_i) \nonumber \\
Y_{ij}\mid\btheta_i &=& f(t_{ij}; \btheta_i) + \gamma_i + \epsilon_{ij}, \ \ (j=1,\ldots,n_i) \label{longresp} \\
\gamma_i &\sim& \mbox{N}(0,\gamma^2) \nonumber \\
\bepsilon_i &\sim& \mbox{MVN}_{n_i}\left( \bzero_{n_i}, \sigma^2 \bR_i \right). \nonumber
\end{eqnarray}
}%
Then, $\phi_i$ corresponds to the probability that patient $i$ has the disease, and the longitudinal trajectory is determined by a parametric function $f(t;\btheta)$.  For instance, for the pregnancy data, we use the sigmoid function
{\small
\begin{equation}
f(t;\btheta) = \frac{\theta_1}{1+\exp\{-\theta_2t - \theta_3\}}  \label{sigmoid},
\end{equation}
}%
which is determined by the parameter vector $\btheta=(\theta_1,\theta_2,\theta_3)$.  The appropriateness of this choice has been well established in the literature ($e.g.$ \cite{MarshallBaronSIM2000, la2006non}). 

The dependence structure of the longitudinal observations is determined by the random intercept $\gamma_i$ and the correlation matrix $\bR_i$.  We assume an autoregressive structure in $\bR_i$ depending on the parameter $\rho$ which represents the correlation between measurements a week apart; that is, the $(j,k)$ element of $\bR_i$ is $\rho^{|t_{ij}-t_{ik}|/7}$.  Hence, corr$(Y_{ij},Y_{ik}) = [\sigma^2\rho^{|t_{ij}-t_{ik}|/7} + \gamma^2 ]/[\sigma^2 + \gamma^2 ]$, which converges to $\gamma^2/[\sigma^2+\gamma^2] > 0$, as $|t_{ij}-t_{ik}|\rightarrow\infty$.  Although it is a common practice to use only an AR structure without a random intercept yielding correlations decreasing toward zero for higher lags, or to use only random intercept with no autoregressive correlation yielding an equicorrelation dependence, including both represents a more realistic scenario. 

Observe that the model is determined by patient-specific parameters $\phi_i$ and $\btheta_i$.  The common two-component model (with a disease-only and a healthy-only cluster) considers two possible values for the parameter vector $(1,\btheta^D)$ and $(0,\btheta^H)$.  
Instead, we propose a Bayesian non-parametric (BNP) model for $(\phi_i,\btheta_i)$ 
in order to obtain clusters of patients with similar longitudinal response trajectories and similar disease statuses; that is, groups of patients with common sets of $(\phi_i,\btheta_i)$.  To this end, we put a Dirichlet Process mixture distribution on the parameter $(\phi_i,\btheta_i)$, such that
{\small
\begin{eqnarray}
(\phi_i,\btheta_i) &\sim& \mathcal{F}, \quad (i=1,\ldots,N) \label{BNP} \\
\mathcal{F} &\sim& \mbox{DirProc}\left(\alpha, \ \mbox{Beta}(a,b)\otimes\mbox{MVN}_3(\btheta^\star, \bSigma) \right), \nonumber
\end{eqnarray}
}%
where the Beta distribution for $\phi_i$ is independent of the multivariate normal for $\btheta_i$ in the base distribution of the random measure $\mathcal{F}$.

Equivalently, we can represent the model (\ref{BNP}) by considering $c_i\in\{1,2,\ldots\}$ to denote the cluster membership of patient $i$, where all patients within the same cluster share the same value for the parameters.  That is, for all $i$ with $c_i=h$, $\phi_i = \phi^{(h)}$ and $\btheta_i = \btheta^{(h)}$.  All patients within a cluster have the same disease probability and the same average longitudinal trajectory.  Using the usual stick-breaking representation \cite{sethuraman1994constructive}, model (\ref{BNP}) is equivalent to
{\small
\begin{eqnarray}
\pr(c_i=h) &=& \psi_h = V_h\prod_{k<h}(1-V_k), \ \ (i=1,\ldots,N) \nonumber \\
V_h &\sim& \mbox{Beta}(1,\alpha), \ \ (h=1,2,\ldots) \label{BNPclust} \\
\phi^{(h)} &\sim& \mbox{Beta}(a,b),\nonumber \\
\btheta^{(h)} &\sim&  \mbox{MVN}_3(\btheta^\star, \bSigma) .\nonumber
\end{eqnarray}
}%
Although the model is specified by an infinite number of clusters $h$, only a finite (often small) number will be realized.  The concentration parameter $\alpha$ helps determine the number of clusters with small $\alpha$ yielding relative few clusters with large memberships and a large $\alpha$ provides many clusters.

In \cite{DelaCruzQuintanaMuller07}, a different BNP specification was previously considered based on the dependent Dirichlet process (DDP; see \cite{deiorio2004}).  This model also induces a clustering of the patients, but the clusters here are designed to contain both diseased and healthy patients and related longitudinal trajectories for each disease status.  In contrast, our approach favors clusters that consist mostly of healthy patients ($\phi^{(h)}$ near zero) or mostly of diseased patients ($\phi^{(h)}$ near one) which facilitates interpretation of results and is more consistent with medical intuition.

Because $\btheta_i$ is a subject-specific parameter vector, one might question the need for the random effect $\gamma_i$ in the response model (\ref{longresp}).  Including this component controls the subject-to-subject variability within clusters, and we have found this reduces the overall number of clusters relative to the model that without this random effect. Further, including $\gamma_i$ in the model leads to fewer clusters with only one or two observations, and the observed clusters tend to have more divergent trajectories $f(t;\btheta^{(k)})$.

In order to classify a new patient according to the disease status based on their (partial) longitudinal trajectory $\by$  observed at time points $\bt$, we compute the probability of disease marginally over their random effect value $\gamma_i$ and their cluster membership $c_i$ using
{\small
\begin{eqnarray}
&&\pr(D=1\,|\,\by) = \sum_{k} \phi^{(k)} \pr(C=k \,|\, \by) \label{predD}\\
&& \ \ \ = \ \frac{\sum_{h} \phi^{(h)}\,\psi_h \exp\left\{ -\frac{1}{2}[\by - f(\bt;\btheta^{(h)}]'[\sigma^2\bR + \gamma^2 \bone\bone']^{-1}[\by - f(\bt;\btheta^{(h)}] \right\}}{\sum_{k} \psi_k \exp\left\{- \frac{1}{2}[\by - f(\bt;\btheta^{(k)}]'[\sigma^2\bR + \gamma^2 \bone\bone']^{-1}[\by - f(\bt;\btheta^{(k)}] \right\}}. \nonumber
\end{eqnarray}
}%

To complete the specification of our Bayesian model, we need prior distributions for the remaining parameters.  For each we choose relatively non-informative priors and make the conditionally conjugate choice when available. We take $\gamma^2 \sim \mbox{InvGamma}(0.1, 0.1)$, $\sigma^2 \sim \mbox{InvGamma}(0.1, 0.1)$, $\rho \sim \mbox{Unif}(0,1)$ for the dependence parameters, and $\alpha\sim \mbox{Gamma}(1,1)$ for the concentration parameter.  For the hyperparameters in the base distributions, we assume $\btheta^\star \sim \mbox{MVN}_3( \bone_3, 10^2\bI_3)$ and $\bSigma \sim \mbox{InvWish}(5, \bI_3)$.  For the Beta distribution of $\phi^{(h)}$, we set the hyperparameters $a=b=0.5$, so that (a priori) we favor values near the extremes.  That is, the cluster should contain mainly healthy or mainly diseased patients.

\section{Model Estimation}\label{estimation}
\subsection{Markov chain Monte Carlo sampling}\label{mcmc}

Posterior inference under this model requires Markov chain Monte Carlo (MCMC) sampling.  To sample from the BNP model, we use the well-known Blackwell-McQueen Polya urn sampler \cite{ishw:jame:2001}.  The BNP model (\ref{BNPclust}) contains an infinite number of potential clusters and must be truncated to $h=1,\ldots,H$ (for some finite $H$) to develop an algorithm.  After selecting the maximum number of clusters $H$, we fix $V_H=1$, so that $\sum_{h=1}^H \psi_h = 1$.  To maintain mixing of the chain, the value of $H$ must be large enough so that there are a number of empty clusters during each iteration.  This protects the sampler from getting trapped in a local mode of the cluster indicators.  In practice, we choose $H$ to be twice the 95th percentile of the number of non-empty clusters, based on a preliminary run of the MCMC chain.

The MCMC algorithm cycles through the following steps during each iteration:
\begin{enumerate}
	\item For $h=1,\ldots,H$, $\phi^{(h)}\sim\mbox{Beta}\left(a + \sum_{i:c_i=h} d_i, b + \sum_{i:c_i=h} (1-d_i) \right)$.
	\item For $h=1,\ldots,H$, sample $\btheta^{(h)}$ from a Metropolis-Hastings (MH) step.  Note that the stationary distribution $p(\btheta^{(h)}\mid\cdots)$ is proportional to
	{\small
	\begin{equation}
	\pi(\btheta^{(h)}\mid\btheta^\star) \prod_{i:c_i=h} \exp\left\{ -\frac{1}{2}[\by_i - f(\bt_i;\btheta^{(h)}]'[\sigma^2\bR_i + \gamma^2 \bone\bone']^{-1}[\by_i - f(\bt_i;\btheta^{(h)}] \right\}.
	\nonumber
	\end{equation}
	}%
	Because the function $f(t;\btheta)$ is not generally factorisable in $\btheta$, we cannot sample the parameter conjugately.  We use an adaptive random-walk MH scheme based on Algorithm \#4 from \cite{andrieu2008}. 
	If the cluster is empty, we instead update $\btheta^{(h)}$ exactly by drawing from the prior $\mbox{MVN}_3(\btheta^\star,\bSigma)$.  
	\item To update $\gamma^2$, $\sigma^2$, and $\rho$, we use Metropolis-Hasting steps.  We propose a candidate value for $\gamma^2$ from $\mbox{logN}(\log\gamma^2,v_\gamma)$ using a psuedo-random walk and consider the posterior likelihood 
	{\small
	\[ p(\gamma^2,\sigma^2,\rho\mid\cdots) \propto \pi(\gamma^2,\sigma^2,\rho) \prod_{i=1}^N p(\by_i\mid c_i, \btheta^{(c_i)}, \gamma^2, \sigma^2, \rho), \]
	}%
	where $\pi(\gamma^2,\sigma^2,\rho)$ is the prior distribution of these correlation parameters and $p(\by_i\mid c_i, \btheta^{(c_i)}, \gamma^2, \sigma^2, \rho)$ is determined by the density of a $\mbox{MVN}_{n_i}( f(\bt_i; \btheta^{(c_i)}), \gamma^2\bone\bone' + \sigma^2\bR_i(\rho))$ distribution evaluated at $\by_i$ and the estimated cluster membership $c_i$. Note that we sample marginally over the $\gamma_i$ random effect.  The move to the proposed $\gamma^2$ is accepted with the usual MH ratio.  Next, a candidate value for $\sigma^2$ is proposed through a psuedo-random walk with a log-Normal candidate distribution and accepted according to the usual MH rules.  Finally, an update to $\rho$ is attempted by a similar random walk MH step.
	\item Update $\alpha\sim\mbox{Gamma}\left(H, 1+\sum_{h=1}^{H-1} \log(1-V_h) \right)$.
	\item For $i=1,\ldots,N$, sample $c_i\sim\mbox{Multinomial}\left(1; (p_1,\ldots,p_H)\right)$, where $p_k = \tilde{p}_k / \sum_{j=1}^H \tilde{p}_j$ and 
	{\small
	\[ \tilde{p}_j = \psi_j(\phi^{(j)})^{d_j}(1-\phi^{(j)})^{1-d_j}\exp\left\{-\frac{1}{2}[\by_i - f(\bt_i;\btheta^{(j)}]'[\sigma^2\bR_i + \gamma^2 \bone\bone']^{-1}[\by_i - f(\bt_i;\btheta^{(j)}] \right\} .\]
	}%
	\item For $h=1,\ldots,H-1$, $V_h \sim \mbox{Beta}(1 + \sum_{i} I(c_i=h),\alpha+\sum_{i} I(c_i>h))$ and set $\psi_h = V_h\prod_{k<h}(1-V_k)$.  Finally, set $\psi_H = \prod_{h=1}^{H-1}(1-V_h)$.
	\item Update $\btheta^\star\sim\mbox{MVN}_3\left(\bm, \bV \right)$, where $\bV = \left[ 10^{-2}\bI_3 + H\bSigma^{-1}\right]^{-1}$ and $\bm = \bV^{-1}\left[ 10^{-2}\bone_3 + \sum_{h=1}^H \bSigma^{-1}\btheta^{(h)} \right]$.
	\item Update $\bSigma\sim\mbox{InvWish}\left(5+H, \bI_3 + \sum_h (\btheta_h - \btheta^\star)(\btheta_h - \btheta^\star)' \right)$.
\end{enumerate}

The MCMC algorithm is run for a large number of iterations to ensure model convergence. 
Due to the large number of parameters and the unidentifiabililty of many, convergence diagnostics are mainly based on two versions of the data likelihood, one conditional on the estimated clusters and the second marginally over the cluster memberships.

\subsection{Inference and model interpretation} \label{clust}

In the context of mixture and BNP models, parameter identifiability and label switching become significant challenges.  This is because the likelihood function is invariant to reparameterization of the cluster labels.  This complicates posterior inference as we cannot simply average parameter values over iterations because clusters are continuously added and deleted over the course of the MCMC chain.  Beyond the hyperparameters ($\alpha,\gamma^2,\sigma^2,\rho,\btheta^\star, \bSigma$), patient-specific predictions are identifiable.  For instance, the predicted disease probability (\ref{predD}) is identified as relabeling the cluster names will not affect this value; the data likelihood (either marginal or conditional on cluster identifications $c_i$) is also identified.  Thus, we can estimate (or predict) the disease status for new patients using Bayesian model averaging (BMA) by averaging the computed values in (\ref{predD}) over all MCMC iterations.

However, estimates of quantities at the cluster level (such as the mean trajectory or probability of disease) cannot be directly obtained from the MCMC posterior sample.  This is a significant challenge as most practitioners, and especially medical clinicians, seek an understanding of the potential trajectories their patients may exhibit and which of these are related to disease.  To accommodate this need, we develop some post-hoc strategies to provide this interpretation.  We address this problem in two stages.  First, we obtain a single point estimate for the clustering of the patients, and secondly, we estimate the disease probabilities and mean HCG trajectory associated with each of the clusters.

We first note that the posterior probability that two patients are assigned to the same cluster, $\pr(C_i=C_j\,|\,y)$, is an identifiable estimand, and we seek to estimate an optimal clustering $\hat{\bc}$.  As the cluster labels are unidentified, we can equivalently call this a partition of the patients.  One commonly used approach due to Dahl (see \cite{dahl2006model}) is to estimate $\hat{\bc}$ by minimizing the loss function
{\small
\begin{equation}
L(\hat{\bc},\bc) = \sum_{i=1}^N \sum_{j=1}^N \left\{ I(\hat{c}_i=\hat{c}_j) - \pr(c_i=c_j\,|\,y) \right\}^2.
\label{dahl}
\end{equation}
}%
Rather than minimizing over the entire space of all possible partitions of $N$ objects, we typically find the partition $\hat{\bc}$ from within the MCMC sample that minimizes the loss.
An alternative approach is to use hierarchical clustering based on the clustering probabilities $\pr(c_i=c_j\,|\,y)$ (see \cite{kaufman2009finding}).  To that end, we define a distance metric $d(i,j) = 1-\pr(c_i=c_j\,|\,y)$ that represents how unlikely patient $i$ and $j$ are to be clustered together, and we create the associated $N\times N$ distance matrix $\bD$ that has $d(i,j)$ as its $(i,j)$th element.  There are a number of different linkage criteria that can be used, but we focus on the average linkage clustering and Ward's Method. These methods produce the entire dendrogram tree, which gives an estimated partition $\hat{\bc}_k$ containing $k$ clusters for each value $k=1,\ldots,N$.

It is then necessary to choose the value of $k$, the number of (non-empty) clusters used to separate the patients.  When using the average linkage method, we may trim the dendrogram at a height $h$ based on the average pairwise clustering probabilities.  That is, for patients $i$ and $j$ in different clusters, $\pr(c_i\neq c_j)$ is on average greater than $h$; we recommend $h=0.75$ or $0.9$ to yield relatively few clusters.  The heights of the dendrogram fit by Ward linkage are not easily interpretable through the pairwise clustering probabilities.  We consider trimming the tree at $k=\hat{K}$, where $\hat{K}$ is the posterior median of the number of non-empty clusters from the MCMC algorithm fit in Section \ref{mcmc}.

There is an ongoing literature on cluster validation methods for hierarchical agglomerative clusterings that also seek to find the appropriate $k$ to use \cite{charrad2014package,Theo:Kout:2008,Mill:Coop:85}.  However, the majority of these approaches require the full data set for validation, as they use quantities such as within and between cluster sum of squares, 
which is not appropriate in our context as our data come from irregular measurement times. 
We consider three validation measures that only use information in the distance matrix $\bD$. 
The Gamma measure \cite{gordon1999} and the Tau measure \cite{milligan81} compare the number of concordant and discordant comparisons.  These comparisons are based of $d(i_1,i_2)$ where $i_1$ and $i_2$ are clustered together in the estimated partition and $d(j_1,j_2)$ where $j_1$ and $j_2$ are in different clusters in the estimated partition.  We also use the Silhouette index \cite{rousseeuw87} which compares the average distance $d(i,j_1)$ across $j_1$ not clustered with $i$ to the average distance $d(i,j_2)$ for $j_2$ clustered with $i$.  For all three, the value of the measure is calculated using the estimated clustering $\hat{\bc}_k$ for each $k$ up to a reasonable upper bound.  The value of $k$ that maximizes the index is chosen to provide the optimal partition.

We select and fix the best clustering $\hat{\bc}$ using either Dahl's method in (\ref{dahl}) or by aggloromative clustering (either average linkage or Ward's method) and selecting the appropriate $k$ ($h$ at fixed value; posterior median $\hat{K}$; Gamma measure; Tau measure; Silhouette index).  Conditional on this clustering, we wish to estimate the parameters associated with these (identified) clusters.  This can be quickly accomplished by obtaining an MCMC sample that does not update the clusterings (skip steps 4.\ and 5.\ in the MCMC algorithm of Section \ref{mcmc}).  This posterior sample is not subject to label switching and provides readily available inference and interpretation.  While providing more interpretable results, this two-stage method will potentially understate the variability due to the unknown number of clusters and the unknown cluster membership.  In the next section, we explore the impact of these features through a simulation study.

\section{Simulation Study}\label{simulations}

\begin{figure}[tb]
\includegraphics[width=\columnwidth]{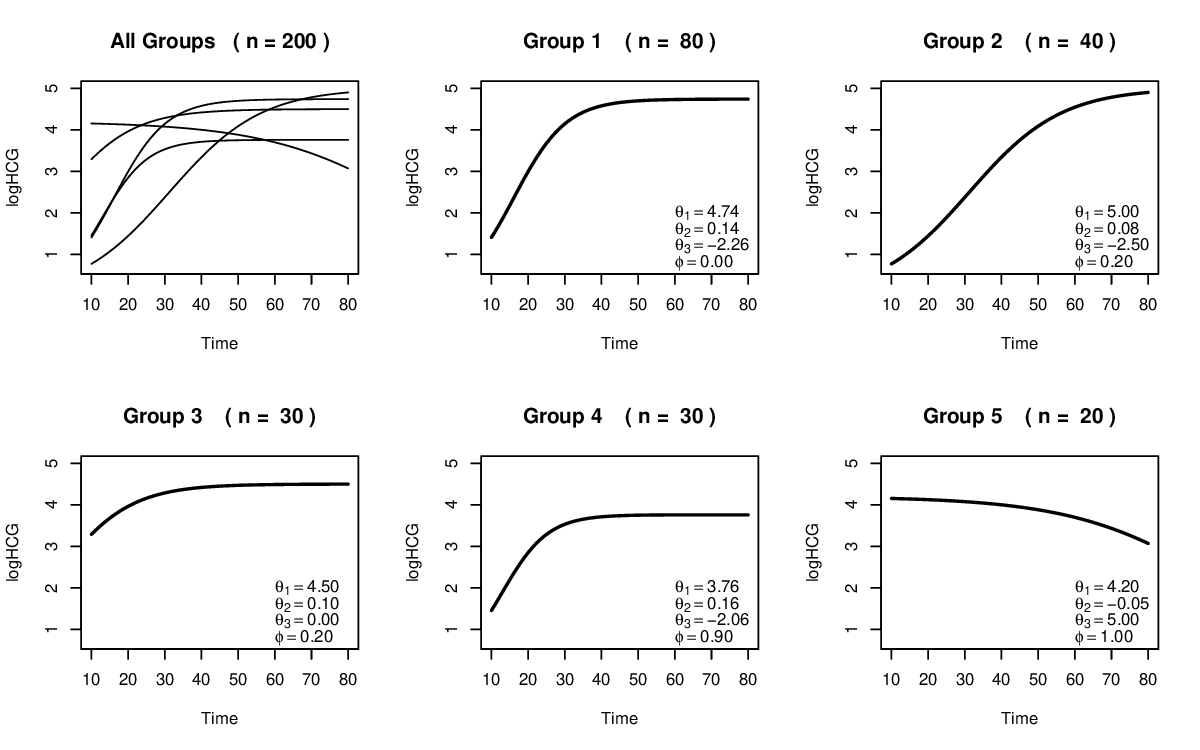}
\caption{Data generating mechanism for Simulation \#1}%
\label{sim1_truth}%
\end{figure}
We consider two situations to assess the accuracy of our methods.  In the first simulation, we consider data that consists of 5 subgroups of patients whose longitudinal trajectories can be seen in Figure \ref{sim1_truth}.  There are 200 patients with 40\%, 20\%, 15\%, 15\%, and 10\% of the sample falling in each cluster, respectively.  The first three clusters represent mainly healthy patients with higher $\beta$--HCG values and lower disease rates of 0\%, 20\%, and 20\%.  The final two groups have lower $\beta$--HCG and higher disease rates (90\% and 100\%).  The true values of the variance parameters are $\sigma^2=0.1$, $\gamma^2=0.05$, and $\rho=0.8$.  We generate 200 datasets each with 200 patients.  We also generate an additional 25,000 patients from the true model to assess out-of-sample predictive accuracy. For comparison, we consider the usual two-component model that has differing trajectory parameters for healthy and diseased patients
{\small
\begin{eqnarray}\label{m2}
D_i &\sim& \mbox{Bern}(\phi) \nonumber \\
Y_{ij}\mid D_i &=& f(t_{ij}; \btheta_{D_i}) + \gamma_i + \epsilon_{ij} \\
\gamma_i\mid D_i &\sim& \mbox{N}(0,\gamma^2_{D_i}) \nonumber \\
\bepsilon_i\mid D_i &\sim& \mbox{MVN}_{n_i}\left( \bzero_{n_i}, \sigma^2_{D_i} \bR_i(\rho_{D_i}) \right), \nonumber
\end{eqnarray}
}%
using prior distributions on the parameters $\phi$, $\btheta_0$, $\btheta_1$, $\gamma^2_d$, $\sigma^2_d$, and $\rho_d$ equivalent to those in the BNP model.  Note that we allow different dependence parameters ($\gamma^2,\sigma^2,\rho$) for healthy patients and diseased patients to increase the flexibility and competitiveness of this choice.  This model is sampled from the MCMC scheme in Section \ref{estimation} with the appropriate modifications.   Note that this scenario should favor our proposed BNP model, as the two-component will fail to account for the heterogeneity within the diseased/healthy populations.

\begin{figure}[tb]
\includegraphics[width=\columnwidth]{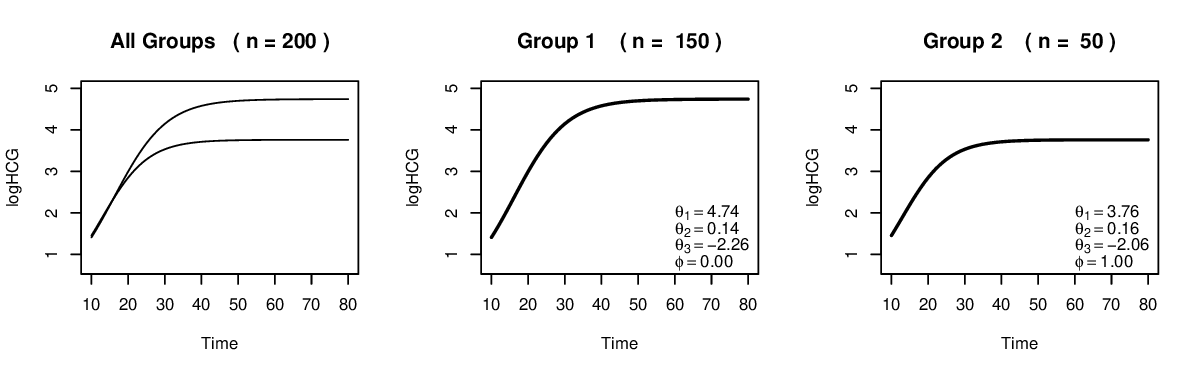}
\caption{Data generating mechanism for Simulation \#2}%
\label{sim2_truth}%
\end{figure}
The second simulation study is designed to favor the two-component model.  The first group with 75\% of patients is disease-free, while group 2 is diseased.   The true values of the variance parameters are $\sigma^2=0.2$, $\gamma^2=0.1$, and $\rho=0.8$. In this setting the two-component model should be optimal, and will help us to assess the impact on prediction accuracy of using the more complex BNP choice. The true longitudinal trajectories are given in Figure \ref{sim2_truth}.  

For each of the 200 datasets in each scenario, we fit the BNP and 2-component models using MCMC.  We then measure the accuracy in prediction using the loss function $L(D,\delta) = N^{-1}\sum_{i} (D_i - \delta_i)^2$, where $D_i$ is the true disease status and $\delta_i= \pr(D_i=1\,|\,Y)$ is the estimated probability of disease as computed by (\ref{predD}).  We  also consider the percent misclassified using $\delta^\star_i = I(\delta_i>1/2)$, the Bayes estimator under the 0-1 loss function, and the area under the curve (AUC) from the receiver operating characteristic (ROC) curve from $\delta_i$ and $D_i$.  Predictions are assessed through within-sample error computed with the 200 patients used to fit the data, as well as out-of-sample accuracy using the model estimates to make predictions for the 25,000 patients in the additional dataset.  Results are contained in Table \ref{sim_results1}.  

\begin{table}[tp]
\caption{Prediction results from simulation studies.  Sample mean and sample standard deviations across the 200 datasets are denoted by $\mathrm{MEAN}_{(\mathrm{SD})}$. The 2-component row gives results from the simple 2-component model, and the BMA row contains prediction results from Bayesian model averaging using the proposed BNP model.  All other rows come from estimating the optimal clustering from the BMA output using various methods, estimating parameters conditional on this clustering, and evaluating the predictions.}
\centering 
\begin{tabular}{l|ccc|ccc}
\hline \hline
& \multicolumn{3}{c|}{Out of Sample} & \multicolumn{3}{c}{Within Sample}\\
Method &  Loss & \% error & AUC & Loss & \% error & AUC \\
\hline
&\multicolumn{3}{c|}{ Simulation Study 1 }&&&\\
2-component  & 0.173$_{(.005)}$ & 25.4\%$_{(0.9)}$ & 0.768$_{(.013)}$ & 0.168$_{(.011)}$ & 25.0\%$_{(2.5)}$ & 0.783$_{(.029)}$ \\
BMA &	 0.151$_{(.003)}$ 	&	 21.7\%$_{(0.7)}$ 	&	 0.823$_{(.007)}$ &	 0.140$_{(.014)}$ 	&	 20.1\%$_{(3.0)}$ 	&	 0.848$_{(.029)}$ 	 \\
Dahl 	&	 0.158$_{(.005)}$ 	&	 22.6\%$_{(0.9)}$ 	&	 0.813$_{(.009)}$ 	&	 0.141$_{(.014)}$ 	&	 20.2\%$_{(2.9)}$ 	&	 0.844$_{(.030)}$ 	 \\
Avg($h=0.75$) 	&	 0.161$_{(.009)}$ 	&	 22.8\%$_{(1.5)}$ 	&	 0.809$_{(.022)}$ 	&	 0.144$_{(.017)}$ 	&	 20.7\%$_{(3.3)}$ 	&	 0.837$_{(.035)}$ 	 \\
Avg($\hat{K}$) &	 0.160$_{(.011)}$ 	&	 22.9\%$_{(1.8)}$ 	&	 0.806$_{(.033)}$ 	&	 0.143$_{(.017)}$ 	&	 20.6\%$_{(3.3)}$ 	&	 0.837$_{(.043)}$ 	 \\
Avg(Silhouette) & 0.162$_{(.007)}$ & 23.0\%$_{(1.1)}$ & 0.806$_{(.018)}$ & 0.147$_{(.016)}$ & 21.2\%$_{(3.2)}$ & 0.831$_{(.036)}$ \\
Avg(Gamma) & 0.160$_{(.011)}$ & 22.8\%$_{(1.9)}$ & 0.806$_{(.035)}$ & 0.143$_{(.019)}$ & 20.5\%$_{(3.6)}$ & 0.834$_{(.043)}$ \\
Avg(Tau) & 0.163$_{(.009)}$ & 23.1\%$_{(1.5)}$ & 0.803$_{(.027)}$ & 0.149$_{(.016)}$ & 21.4\%$_{(3.2)}$ & 0.827$_{(.037)}$ \\
Ward($\hat{K}$)	 &	 0.160$_{(.012)}$ 	&	 22.9\%$_{(2.1)}$ 	&	 0.806$_{(.036)}$ 	&	 0.144$_{(.019)}$ 	&	 20.7\%$_{(3.6)}$ 	&	 0.836$_{(.043)}$ 	 \\
Ward(Silhouette) & 0.164$_{(.013)}$ & 23.4\%$_{(2.2)}$ & 0.800$_{(.036)}$ & 0.149$_{(.018)}$ & 21.6\%$_{(3.5)}$ & 0.825$_{(.048)}$ \\
Ward(Gamma) & 0.160$_{(.013)}$ & 23.0\%$_{(2.2)}$ & 0.804$_{(.043)}$ & 0.144$_{(.021)}$ & 20.7\%$_{(3.9)}$ & 0.835$_{(.052)}$ \\ 
Ward(Tau) & 0.162$_{(.010)}$ & 23.1\%$_{(1.7)}$ & 0.802$_{(.033)}$ & 0.148$_{(.018)}$ & 21.5\%$_{(3.5)}$ & 0.828$_{(.043)}$ \\
\hline
&\multicolumn{3}{c|}{ Simulation Study 2 }&&&\\
2-component & 0.132$_{(.008)}$ & 18.2\%$_{(1.0)}$ & 0.853$_{(.007)}$ & 0.114$_{(.011)}$ & 16.0\%$_{(2.1)}$ & 0.877$_{(.027)}$ \\
 BMA 	&	 0.128$_{(.007)}$ 	&	 18.0\%$_{(1.0)}$ 	&	 0.854$_{(.008)}$ 	&	 0.115$_{(.012)}$ 	&	 16.0\%$_{(2.0)}$ 	&	 0.875$_{(.028)}$ 	 \\
 Dahl 		&	 0.130$_{(.008)}$ 	&	 18.0\%$_{(1.0)}$ 	&	 0.856$_{(.008)}$ 	&	 0.114$_{(.012)}$ 	&	 16.1\%$_{(2.2)}$ 	&	 0.875$_{(.027)}$ 	 \\
 Avg($h=0.75$) 		&	 0.130$_{(.008)}$ 	&	 18.0\%$_{(1.1)}$ 	&	 0.857$_{(.009)}$ 	&	 0.115$_{(.012)}$ 	&	 16.1\%$_{(2.0)}$ 	&	 0.874$_{(.028)}$ 	 \\
 Avg($\hat{K}$) 		&	 0.132$_{(.014)}$ 	&	 18.2\%$_{(1.6)}$ 	&	 0.851$_{(.032)}$ 	&	 0.116$_{(.018)}$ 	&	  16.3\%$_{(2.7)}$ 	&	 0.871$_{(.045)}$ 	 \\
Avg(Silhouette) & 0.130$_{(.008)}$ & 18.0\%$_{(1.1)}$ & 0.856$_{(.009)}$ & 0.115$_{(.012)}$ & 16.1\%$_{(2.2)}$ & 0.874$_{(.029)}$ \\
Avg(Gamma) & 0.131$_{(.008)}$ & 18.1\%$_{(1.0)}$ & 0.851$_{(.013)}$ & 0.115$_{(.012)}$ & 16.2\%$_{(2.1)}$ & 0.874$_{(.028)}$ \\
Avg(Tau)  & 0.135$_{(.012)}$ & 18.9\%$_{(1.7)}$ & 0.840$_{(.021)}$ & 0.118$_{(.015)}$ & 16.4\%$_{(2.6)}$ & 0.869$_{(.038)}$ \\
 Ward($\hat{K}$) 	 	&	 0.132$_{(.011)}$ 	&	 18.2\%$_{(1.2)}$ 	&	 0.852$_{(.023)}$ 	&	 0.116$_{(.014)}$ 	&	 16.2\%$_{(2.3)}$ 	&	 0.872$_{(.036)}$ 	 \\
Ward(Silhouette) & 0.130$_{(.008)}$ & 18.0\%$_{(1.0)}$ & 0.857$_{(.009)}$ & 0.115$_{(.012)}$ & 16.1\%$_{(2.1)}$ & 0.874$_{(.028)}$ \\
Ward(Gamma)  & 0.131$_{(.011)}$ & 18.1\%$_{(1.1)}$ & 0.853$_{(.012)}$ & 0.115$_{(.014)}$ & 16.1\%$_{(2.2)}$ & 0.874$_{(.028)}$ \\
Ward(Tau)  & 0.132$_{(.009)}$ & 18.4\%$_{(1.5)}$ & 0.846$_{(.018)}$ & 0.117$_{(.013)}$ & 16.3\%$_{(2.4)}$ & 0.871$_{(.032)}$ \\
\hline\hline
\end{tabular}
\label{sim_results1}
\end{table}

In the first simulation study, it is clear that the BNP model dominates the 2-component choice.  Both the out-of- and within-sample estimation risks are around 15\% higher for the two-component model compared to the predictions from the model averaged BNP choice.  Treating all diseased/healthy patients as coming from the same model leads to poor prediction when this is not the true model.  
As described in Section \ref{clust}, after fitting the full BNP model, we use a variety of methods to estimate the best clustering of the patients and re-estimate the posterior parameters conditional on this $\hat{\bc}$.  All methods that fail to mix over the unknown $c_i$ suffer from 5--8\% higher out-of-sample loss relative the model averaged predictions; within-sample accuracy is much less impacted.  While Dahl's method has one of the better accuracies of the second stage estimators, there is relatively little difference across the choices of estimating $\hat{\bc}$.

In the second simulation, the true model is the 2-component model.  However, the model averaged predictions from the more complex BNP are equivalent or slightly better than the true model for out-of-sample predictions.  The within-sample predictions are similar.  Prediction accuracy is similar across all the stage 2 estimators, with the possible exception of the methods the determine the number of clusters using the Tau criteria.

\begin{table}[tp]
\caption{Clustering results from simulation studies.  For each simulation study, we display the number of non-empty clusters, the geometric mean cluster size, the number of small clusters (fewer than 5 patients), and the partition error.}
\centering 
\begin{tabular}{l|cccc|cccc}
\hline \hline
& \multicolumn{4}{c|}{Simulation Study 1} & \multicolumn{4}{c}{Simulation Study 2}\\
\multirow{2}{*}{Method} & \# of & Avg. 	& Small 	& \multirow{2}{*}{Error} 	&\# of & Avg. 	& Small 	& \multirow{2}{*}{Error} 	\\
						& Clusters & Size 	& Clusters	&							& Clusters & Size 	& Clusters	&							\\
\hline
Truth	&	5			&	35.7			&	0			&				&		2			&	87			&	0			&				 \\					
2-component 	&	2$_{(	0	)}$ &	92$_{(	2	)}$ &	0.0$_{(	0	)}$ &	0.37$_{(	0.01	)}$ &		2$_{(	0	)}$ &	87$_{(	0	)}$ &	0$_{(	0	)}$ &	0$_{(	0	)}$ \\					
BMA 	&	8.1$_{(	1.1	)}$ &	16$_{(	3	)}$ &	2.5$_{(	0.6	)}$ &	0.18$_{(	0.02	)}$ &		3.0$_{(	0.6	)}$ &	56$_{(	11	)}$ &	0.5$_{(	0.3	)}$ &	0.14$_{(	0.08	)}$ \\					
Dahl 	&	9.0$_{(	2.9	)}$ &	13$_{(	8	)}$ &	3.2$_{(	2.4	)}$ &	0.16$_{(	0.02	)}$ &		2.5$_{(	0.8	)}$ &	65$_{(	30	)}$ &	0.3$_{(	0.6	)}$ &	0.04$_{(	0.08)}$ \\					
Avg($h=0.75$) 	&	5.0$_{(	1.1	)}$ &	31$_{(	11	)}$ &	0.5$_{(	0.9	)}$ &	0.15$_{(	0.02)}$ &		2.1$_{(	0.2	)}$ &	84$_{(	12	)}$ &	0.0$_{(	0.2	)}$ &	0.02$_{(	0.05)}$ \\
Avg($\hat{K}$) 	&	7.7$_{(	1.1	)}$ &	10$_{(	3	)}$ &	2.9$_{(	1.0	)}$ &	0.15$_{(	0.02)}$ &		2.7$_{(	0.7	)}$ &	57$_{(	31	)}$ &	0.4$_{(	0.6	)}$ &	0.04$_{(	0.08)}$ \\
Avg(Silhouette) 	&	4.2$_{(	0.9	)}$ &	41$_{(	10	)}$ &	0.1$_{(	0.5	)}$ &	0.16$_{(	0.02	)}$ &		2.1$_{(	0.4	)}$ &	84$_{(	11	)}$ &	0.0$_{(	0.1	)}$ &	0.03$_{(	0.07)}$ \\
Avg(Gamma) 	&	10.7$_{(	2.2	)}$ &	7$_{(	7	)}$ &	5.8$_{(	2.1	)}$ &	0.15$_{(	0.02	)}$ &		4.7$_{(	3.7	)}$ &	48$_{(	39	)}$ &	2.4$_{(	3.4	)}$ &	0.05$_{(	0.09	)}$ \\					
Avg(Tau) 	&	4.1$_{(	0.5	)}$ &	42$_{(	9	)}$ &	0.1$_{(	0.3	)}$ &	0.16$_{(	0.02	)}$ &		9.8$_{(	3.4	)}$ &	9$_{(	17	)}$ &	7.4$_{(	3.7	)}$ &	0.08$_{(	0.07	)}$ \\					
Ward($\hat{K}$)	&	7.7$_{(	1.1	)}$ &	10$_{(	3	)}$ &	2.9$_{(	1.0	)}$ &	0.15$_{(	0.02	)}$ &		2.7$_{(	0.7	)}$ &	57$_{(	31	)}$ &	0.4$_{(	0.6	)}$ &	0.04$_{(	0.08)}$ \\
Ward(Silhouette) 	&	4.1$_{(	0.6	)}$ &	44$_{(	8	)}$ &	0.0$_{(	0.1	)}$ &	0.16$_{(	0.02	)}$ &		2.1$_{(	0.3	)}$ &	85$_{(	9	)}$ &	0.0$_{(	0.1	)}$ &	0.03$_{(	0.07)}$ \\
Ward(Gamma) 	&	8.2$_{(	3.0	)}$ &	19$_{(	11	)}$ &	2.1$_{(	2.3	)}$ &	0.15$_{(	0.02	)}$ &		3.0$_{(	2.0	)}$ &	66$_{(	30	)}$ &	0.6$_{(	1.5	)}$ &	0.05$_{(	0.09)}$ \\
Ward(Tau) 	&	4.2$_{(	0.7	)}$ &	42$_{(	9	)}$ &	0.0$_{(	0.1	)}$ &	0.16$_{(	0.02	)}$ &		5.7$_{(	4.0	)}$ &	38$_{(	34	)}$ &	3.0$_{(	4.1	)}$ &	0.08$_{(	0.08)}$ \\					
\hline\hline
\end{tabular}
\label{sim_results2}
\end{table}
We further explore the behavior of the estimated partitions in Table \ref{sim_results2}.  For each simulation and model fit, we display the average number of non-empty clusters fit to each method.  The BMA results are the number of non-empty clusters averaged across MCMC iterations, averaged across datasets.  We also consider two measures related to the size of the estimated clusters:  the geometric mean cluster size (arithmetic mean is necessarily the inverse of the number of clusters), and the number of ``small'' clusters, defined to contain five or fewer members.  To ease interpretation, one would prefer fewer, more dense clusters.  Finally, we compare the estimated cluster membership to the truth by considering the average of $|I(\hat{c}_i=\hat{c}_j) - I(c_i=c_j)|$ over all $(i,j)$ pairs of patients.  In the BMA, we replace $I(\hat{c}_i=\hat{c}_j)$ with estimated posterior matching probability.
 
In the first simulation study, we recognize that the BMA methods tends to overestimated the number of clusters at $8.1\pm 1.1$, relative to the truth equal to  5.  In the second study, there is only a slight tendency toward extraneous clusters, but as noted previously, this helps avoid overly confident predictions.  Looking toward the estimated clustering $\hat{\bc}$, the Dahl method often produces more clusters than are necessary.  In the first simulation, this method also had much larger variability across data sets (range: 4--19).  Considering the hierarchical clustering methods, using the average versus the Ward linkage function produces similar groupings within each method for choosing the number of clusters.  The Silhouette approach tends to produce fewer clusters, relative to the Gamma method.  The Tau criteria performs poorly, selecting too few clusters  in the first setting and  too many in the second.  The subjective choice of cutting the average linkage dendrogram at the height of $h=0.75$ produces favorable results.

In conclusion, we recommend the use the model averaged results for prediction when possible.  To make cluster-specific interpretation, the simulation study indicates hierarchical clustering with the average linkage cut at $h=0.75$ performs well, along with the Silhouette criteria under either linkage and and the Dahl method.  

\section{Pregnancy Data Example}\label{application}
We apply our methodology to the Chilean assisted reproductive therapy (ART) data introduced in Section \ref{data}.  The goal is to predict the outcome of an ART-induced pregnancy (whether the baby is born alive) based on the longitudinal $\beta$--HCG measurements.  To this end, we fit the proposed BNP model (with the different estimation methods) and the two-component model defined in (\ref{m2}) for comparison.  Note that we now only consider the four choices for determining $\hat{\bc}$ that performed best in the simulations.   We use the full data with $N=173$ patients and assess the within-sample error using the same metrics as in the simulation study: the loss function $L(D,\delta) = N^{-1}\sum_{i} (D_i - \delta_i)^2$, the percent misclassified by $\delta_i^\star = I(\delta_i>1/2)$, and the AUC from the ROC curve.  To approximate the out-of-sample error, we perform 25-fold cross validation (CV) by withholding a random 35 patients (about 20\% of the data) from model estimation to use as test data.  

Estimation performance is summarized in Table \ref{application_results}.  
\begin{table}[t]
\caption{Prediction accuracy in Chilean women application}
\centering
\begin{tabular}{l|cccc|ccc}
\hline \hline
& \multicolumn{4}{c|}{Full data} & \multicolumn{3}{c}{25-fold CV ($n_\text{test}=35; n_\text{train}=138$)}\\
Method & Clusters & Loss & \% error & AUC & Loss & \% error & AUC \\
\hline
2-component & 2 & 0.124 & 16.2\% & 0.863 & 0.141$_{(.008)}$ & 19.5\%$_{(1.2)}$ & 0.865$_{(.004)}$ \\
BMA & 8.6 & 0.099 & 13.3\% & 0.900 & 0.124$_{(.007)}$ & 16.2\%$_{(1.1)}$& 0.900$_{(.003)}$\\
Dahl & 10 & 0.099 & 12.7\% & 0.898  & 0.129$_{(.007)}$& 16.8\%$_{(1.0)}$& 0.888$_{(.004)}$\\
Avg($h=0.75$) & 5 & 0.103 & 13.3\% & 0.883 & 0.129$_{(.008)}$ & 17.2\%$_{(1.4)}$ & 0.881$_{(.005)}$ \\
Avg(Silhouette) & 5 & 0.100 & 13.3\% & 0.880 & 0.126$_{(.007)}$ & 16.3\%$_{(1.0)}$ & 0.890$_{(.003)}$ \\
Ward(Silhouette) & 3 & 0.104 & 14.5\% & 0.889 & 0.131$_{(.008)}$ & 17.0\%$_{(1.0)}$ & 0.884$_{(.005)}$ \\
\hline \hline
\end{tabular}
\label{application_results}
\end{table}
We observe the two-component model does significantly worse than the BNP approaches.  Relative to the Bayes model averaged BNP approach, this naive classifier has 25\% higher within-sample loss and 15\% higher out-of-sample loss. The percent misclassified and the AUC are similarly worse.  Figure {\color{blue} 4} 
displays the estimated probabilities of complication (based on the full data analysis) as computed by (\ref{predD}) under BMA.  Additionally, we compute a 50\% credible interval for $\pr(D_i=1\mid\by)$ for each patient based on the quantiles of the per-iteration estimates.  As can be seen in the figure, classification based on $I(\delta_i>1/2)$ may not be well-calibrated (98\% specificity but only 57\% sensitivity), possibly due to the fact that only 28\% of patients experienced complications.  Using the ROC curve to find the best threshold assuming an equal cost of Type I and Type II errors, the test based on $I(\delta_i>0.23)$ has 90\% specificity and 80\% sensitivity.  This threshold could also be computed based on the recognition that a false negative is more costly than a false positive \cite{Perk:Schi:2006}.

\begin{figure}[t]
\includegraphics[width= 1\textwidth]{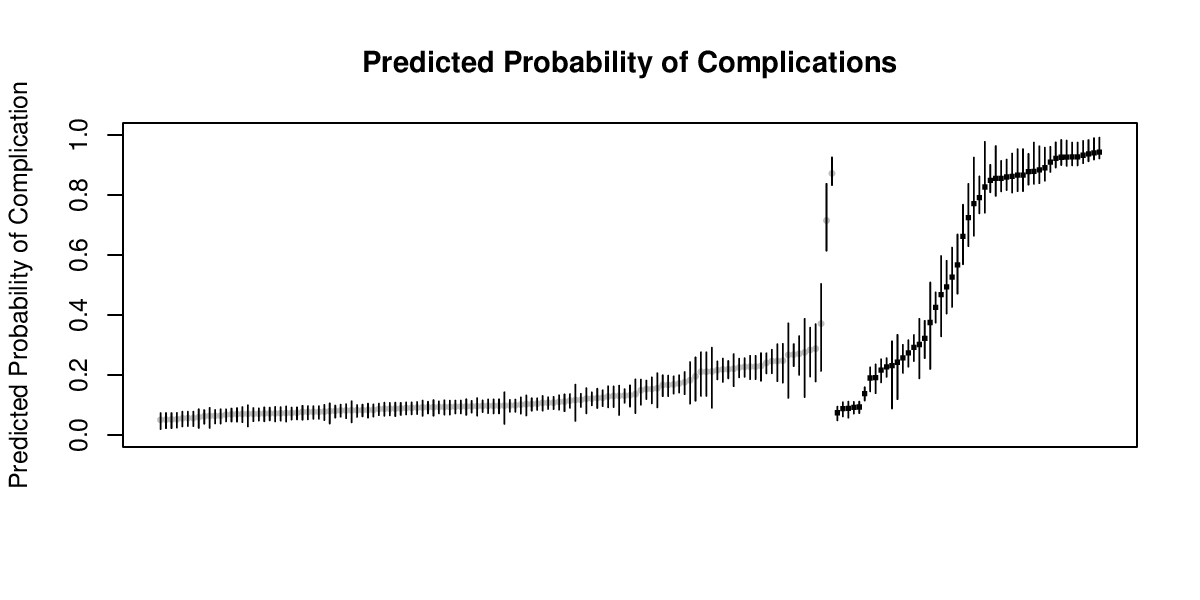}
\label{predictions}
\caption{Estimated probability of complication in the Chilean pregnancy data, along with a 50\% credible interval.  Patients depicted by a gray dot had no complication, whiles those  denoted by a black square did.}
\end{figure}

We next compare the predictions and the estimates from the various BNP models.  All methods perform failry well for the in-sample predictions.  For the cross-validations results, the BMA obviously is strongest, followed by the hierarchical clustering with the thresholds chosen by the Silhouette criterion.  We also note that Dahl's approach does reasonably well in terms of prediction accuracy, but it estimates an optimal partition with 10 clusters, much more than agglomerative clustering methods.  This is a significant drawback for interpretation with no appreciable benefit in terms of prediction accuracy.  To provide interpretable results describing the impact of $\beta$--HCG trajectory on pregnancy complications, we focus on the results from the Avg(Silhouette) analysis.

\begin{figure}[t]
\includegraphics[width= 1\textwidth]{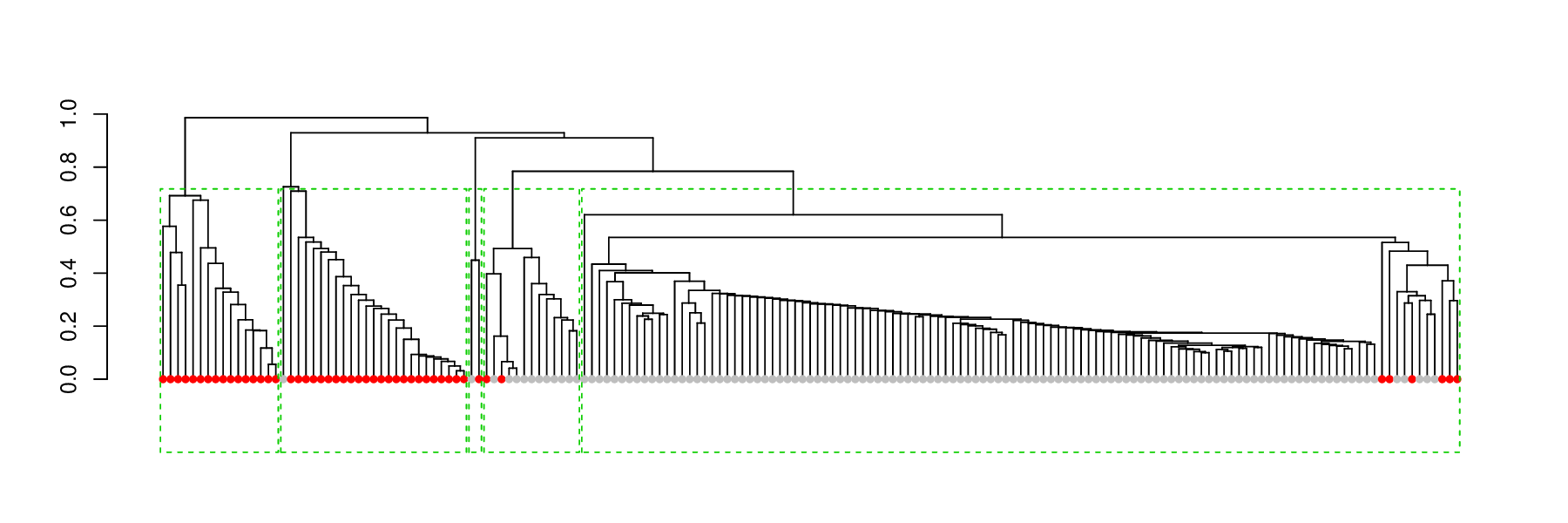}
\label{dendro}
\caption{Dendrogram describing the hierarchical clustering structure with average linkage function of patients in Chilean pregnancy data.  Grey circles indicate patients with no complications, and red circles for those with complications.  The dashed line boxes display the Silhouette estimated partition.}
\end{figure}
Figure {\color{blue}5} 
  displays the dendrogram depicting the relationship across patients determined by the hierarchical agglomerative clustering method with the average linkage function.  Based on the Silhouette criteria, we select the partition with $K=5$ clusters.  
\begin{figure}[t]
\includegraphics[width= 1\textwidth]{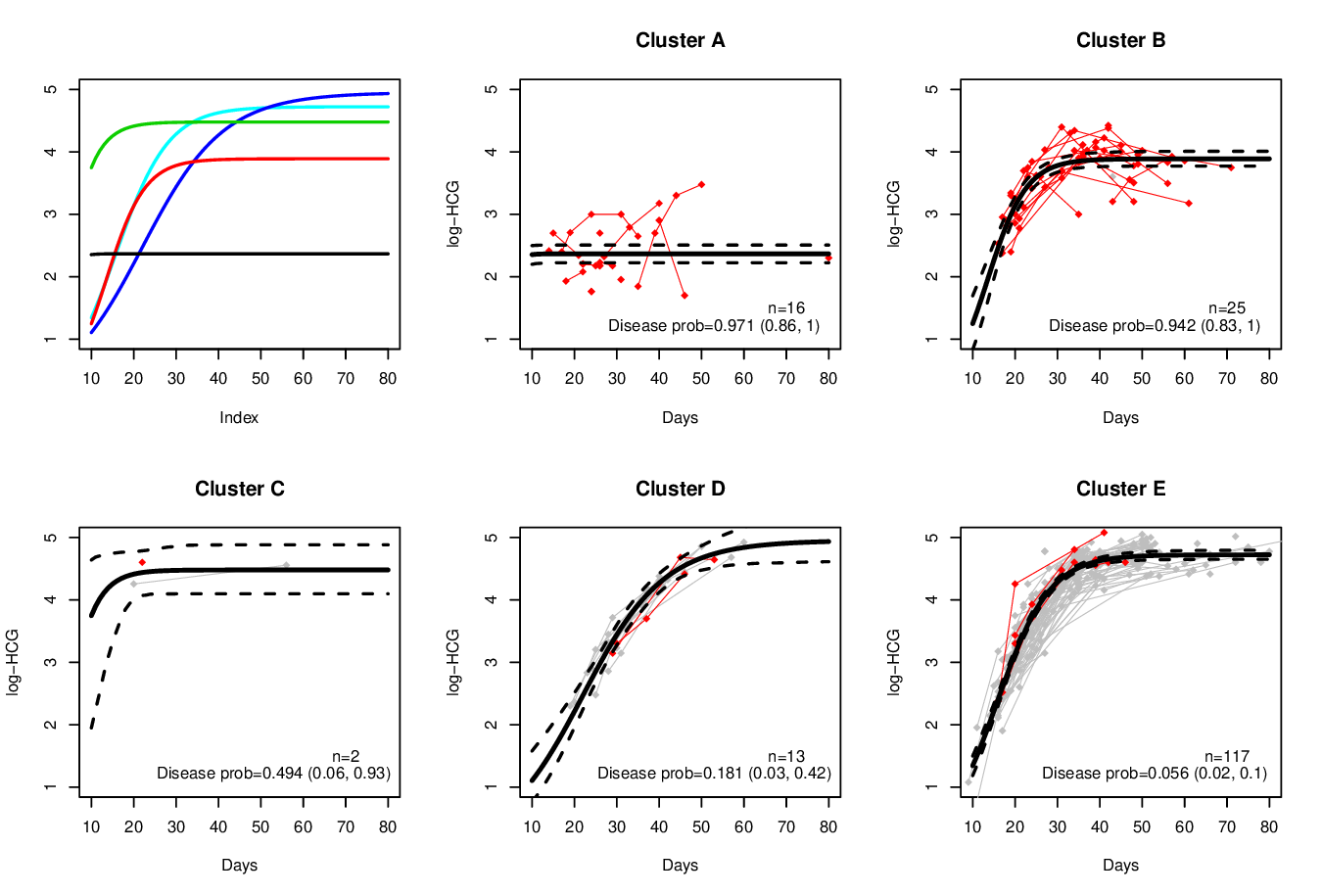}
\label{traj}
\caption{Estimated trajectories for the five clusters estimated by the hierarchical clustering model.}
\end{figure}
Figure {\color{blue} 6} 
displays the trajectories for each of these five clusters, along with the spaghetti plots of the patients assigned to each.  This represents information that can easily be communicated to an obstetrician to inform their practice.
For instance, it is immediately clear the cluster E captures that majority of patients and represents those with a low chance of a complication.  These healthy patients experience rapidly increasing $\beta$--HCG between days 10 and 30 and stabilize at a value near 4.7.  Clusters A and B both represent patients who experience complications, but they do so for two different reasons.  Cluster A patients never experience an increase in HCG.  Conversely, patients in cluster B show the usual rapid increase (as in cluster E) but their levels plateau too early and at a lower height of around 3.9.  These two clusters clearly represent different pathologies with different $\beta$--HCG trajectories.  The usual two-component model (as shown in Figure \ref{spaghetti}) 
 estimates a single model here and cannot distinguish between these two trajectories.  Relative to cluster E, cluster D represents patients who are at an elevated chance of complication while still predicted to be healthy; these patients show a slower increase in HCG but eventually reach a similar level.  While we would still predict a woman in cluster D to have a successful pregnancy, these patients might be flagged for closer follow-up than patients in cluster E.  Cluster C represents 2 outlier patients, one healthy and one diseased, with abnormally high $\beta$--HCG early in the trimester.  With a 50\% disease rate and small size, this cluster does not provide much clinical information.


\section{Discussion}\label{disc} 
Motivated by a study looking at the longitudinal profiles of the $\beta$--HCG hormone levels from a pregnant women in Chile, this manuscript considers the general problem  of classification based on longitudinal trajectories when no other covariates are available. In this context we introduced a Bayesian non-parametric model that increases the flexibility of the standard semi-parametric mixed-effect models beyond the usual two-component choice. 
One of the key aspects of our procedure is that we automatically allow multiple trajectories/pathologies to be associated with the diseased and healthy populations.  This is facilitates interpretation and is more biologically plausible than the two-component choice.  Prediction can be performed using Bayesian model averaging, and our proposed strategy for selecting the optimal number of clusters (when this number is unknown) facilitates model interpretation. Simulation experiments and the real data application illustrate the advantages of the proposed procedure and demonstrate that inference conditional on the estimated clustering $\hat{\bc}$ did not lead to any significant loss of accuracy.

Prediction of disease using both the longitudinal profile and additional covariates provides a natural extension of this model.  For instance, the age of the mother and the use of frozen vs.\ fresh embryo are important factors that could be taken into account.  The data analyzed in this paper did not include any covariate information, but the modeling framework could be appropriately extended.  This is left for future research.

Implementation of the model can be accomplished following the MCMC algorithm outlined in this paper. Software is available from the authors upon request.

\bibliographystyle{acmtrans-ims}
\bibliography{BNPclass}

\begin{thebibliography}{}
\ifx \url   \undefined \def \url#1{#1}   \fi

\bibitem{andrieu2008}
\textsc{Andrieu, C.} \textsc{and} \textsc{Thoms, J.} (2008).
\newblock A tutorial on adaptive {MCMC}.
\newblock \emph{Statistics and Computing\/}~\textbf{18},~4, 343--373.

\bibitem{arribas2015classification}
\textsc{Arribas-Gil, A.}, \textsc{De~la Cruz, R.}, \textsc{Lebarbier, E.},
  \textsc{and} \textsc{Meza, C.} (2015).
\newblock Classification of longitudinal data through a semiparametric
  mixed-effects model based on lasso-type estimators.
\newblock \emph{Biometrics\/}~\textbf{71},~2, 333--343.

\bibitem{charrad2014package}
\textsc{Charrad, M.}, \textsc{Ghazzali, N.}, \textsc{Boiteau, V.},
  \textsc{Niknafs, A.}, \textsc{and} \textsc{Charrad, M.~M.} (2014).
\newblock Package �nbclust�.
\newblock \emph{J. Stat. Soft\/}~\emph{61}, 1--36.

\bibitem{obs2}
\textsc{Confino, E.}, \textsc{Demir, R.~H.}, \textsc{Friberg, J.}, \textsc{and}
  \textsc{Gleicher, N.} (1986).
\newblock The predictive value of hcg beta subunit levels in pregnancies
  achieved by in vitro fertilization and embryo transfer: an international
  collaborative study.
\newblock \emph{Fertility and Sterility\/}~\emph{45}, 526--531.

\bibitem{dahl2006model}
\textsc{Dahl, D.~B.} (2006).
\newblock Model-based clustering for expression data via a dirichlet process
  mixture model.
\newblock \emph{Bayesian inference for gene expression and proteomics\/},
  201--218.

\bibitem{deiorio2004}
\textsc{De~Iorio, M.}, \textsc{M\"{u}ller, P.}, \textsc{Rosner, G.~L.},
  \textsc{and} \textsc{MacEachern, S.~N.} (2004).
\newblock An {ANOVA} model for dependent random measurers.
\newblock \emph{Journal of the American Statistical
  Association\/}~\textbf{99},~465, 205--215.

\bibitem{de2017predicting}
\textsc{De~la Cruz, R.}, \textsc{Fuentes, C.}, \textsc{Meza, C.}, \textsc{Lee,
  D.-J.}, \textsc{and} \textsc{Arribas-Gil, A.} (2017).
\newblock Predicting pregnancy outcomes using longitudinal information: a
  penalized splines mixed-effects model approach.
\newblock \emph{Statistics in Medicine\/}~\textbf{36},~13, 2120--2134.

\bibitem{de2016error}
\textsc{de~la Cruz, R.}, \textsc{Fuentes, C.}, \textsc{Meza, C.}, \textsc{and}
  \textsc{N{\'u}{\~n}ez-Ant{\'o}n, V.} (2016).
\newblock Error-rate estimation in discriminant analysis of non-linear
  longitudinal data: A comparison of resampling methods.
\newblock \emph{Statistical methods in medical research\/},
  https://doi.org/10.1177/0962280216656246.

\bibitem{de2016bayesian}
\textsc{De~la Cruz, R.}, \textsc{Meza, C.}, \textsc{Arribas-Gil, A.},
  \textsc{and} \textsc{Carroll, R.~J.} (2016).
\newblock Bayesian regression analysis of data with random effects covariates
  from nonlinear longitudinal measurements.
\newblock \emph{Journal of multivariate analysis\/}~\emph{143}, 94--106.

\bibitem{DelaCruzQuintana07}
\textsc{De~la Cruz-Mes\'{\i}a, R.} \textsc{and} \textsc{Quintana, F.~A.}
  (2007).
\newblock A model--based approach to {B}ayesian classification with
  applications to predicting pregnancy outcomes from longitudinal
  $\beta$--h{CG} profiles.
\newblock \emph{Biostatistics\/}~\emph{8}, 228--238.

\bibitem{DelaCruzQuintanaMuller07}
\textsc{De~la Cruz-Mes\'{\i}a, R.}, \textsc{Quintana, F.~A.}, \textsc{and}
  \textsc{M\"uller, P.} (2007).
\newblock Semiparametric {B}ayesian classification with longitudinal markers.
\newblock \emph{Journal of the Royal Statistical Society, Series
  C\/}~\emph{56}, 119--137.

\bibitem{obs3}
\textsc{Frits, M.~A.} \textsc{and} \textsc{Guo, S.~M.} (1987).
\newblock Doubling time of human chorionic gonadotropin (hcg) in early normal
  pregnancy: relationship to hcg concentration and gestational age.
\newblock \emph{Fertility and Sterility\/}~\emph{47}, 584--589.

\bibitem{gershman2012}
\textsc{Gershman, S.~J.} \textsc{and} \textsc{Blei, D.~M.} (2012).
\newblock A tutorial on {B}ayesian nonparametric models.
\newblock \emph{Journal of Mathematical Psychology\/}~\textbf{56},~1, 1--12.

\bibitem{gordon1999}
\textsc{Gordon, A.~D.} (1999).
\newblock \emph{Classification}, 2 ed.
\newblock Chapman \& Hall/CRC.

\bibitem{hjort2010bayesian}
\textsc{Hjort, N.~L.}, \textsc{Holmes, C.}, \textsc{M{\"u}ller, P.},
  \textsc{and} \textsc{Walker, S.~G.} (2010).
\newblock \emph{Bayesian nonparametrics}.  Vol.~\textbf{28}.
\newblock Cambridge University Press.

\bibitem{ishw:jame:2001}
\textsc{Ishwaran, H.} \textsc{and} \textsc{James, L.~F.} (2001).
\newblock Gibbs sampling methods for stick-breaking priors.
\newblock \emph{Journal of the American Statistical
  Association\/}~\textbf{96},~453, 161--173.

\bibitem{kaufman2009finding}
\textsc{Kaufman, L.} \textsc{and} \textsc{Rousseeuw, P.~J.} (2009).
\newblock \emph{Finding groups in data: an introduction to cluster analysis}.
  Vol.~\textbf{344}.
\newblock John Wiley \& Sons.

\bibitem{la2006non}
\textsc{la~Cruz-Mes{\'\i}a, D.}, \textsc{Marshall, G.}, \textsc{and}
  \textsc{others}. (2006).
\newblock Non-linear random effects models with continuous time autoregressive
  errors: a bayesian approach.
\newblock \emph{Statistics in medicine\/}~\textbf{25},~9, 1471--1484.

\bibitem{lutsetal:12}
\textsc{Luts, J.}, \textsc{Molenberghs, G.}, \textsc{Verbeke, G.},
  \textsc{Van~Huffel, S.}, \textsc{and} \textsc{Suykens, J. A.~K.} (2012).
\newblock A mixed effects least squares support vector machine model for
  classification of longitudinal data.
\newblock \emph{Computational Statistics and Data Analysis\/}~\emph{56},
  611--628.

\bibitem{MarshallBaronSIM2000}
\textsc{Marshall, G.} \textsc{and} \textsc{Bar\'on, A.~E.} (2000).
\newblock Linear discriminant models for unbalanced longitudinal data.
\newblock \emph{Statistics in Medicine\/}~\emph{19}, 1969--1981.

\bibitem{milligan81}
\textsc{Milligan, G.~W.} (1981).
\newblock A {M}onte {C}arlo study of thirty internal criterion measures for
  cluster analysis.
\newblock \emph{Psychometrika\/}~\textbf{45},~3, 325--342.

\bibitem{Mill:Coop:85}
\textsc{Milligan, G.~W.} \textsc{and} \textsc{Cooper, M.~C.} (1985).
\newblock An examination of procedures for determining the number of clusters
  in a data set.
\newblock \emph{Psychometrika\/}~\emph{50(2)}, 159--179.

\bibitem{muller:05}
\textsc{M\"{u}ller, H.-G.} (2005).
\newblock Functional modelling and classification of longitudinal data.
\newblock \emph{Scandinavian Journal of Statistics\/}~\emph{32}, 223--240.

\bibitem{Perk:Schi:2006}
\textsc{Perkins, N.~J.} \textsc{and} \textsc{Schisterman, E.~F.} (2006).
\newblock The inconsistency of ``optimal'' cutpoints obtained using two
  criteria based on the receiver operating characteristic curve.
\newblock \emph{American Journal of Epidemiology\/}~\emph{167(7)}, 670--675.

\bibitem{rousseeuw87}
\textsc{Rousseeuw, P.} (1987).
\newblock Silhouettes: {A} graphical aid to the interpretation and validation
  of cluster analysis.
\newblock \emph{Journal of Computational and Applied Mathematics\/}~\emph{20},
  53--65.

\bibitem{sethuraman1994constructive}
\textsc{Sethuraman, J.} (1994).
\newblock A constructive definition of dirichlet priors.
\newblock \emph{Statistica sinica\/}, 639--650.

\bibitem{obs1}
\textsc{Shepherd, R.~W.}, \textsc{Patton, P.~E.}, \textsc{Novy, M.~J.},
  \textsc{and} \textsc{A., B.~K.} (1990).
\newblock Serial beta-hcg measurements in the early detection of ectopic
  pregnancy.
\newblock \emph{Obstetrics and Gynecology\/}~\emph{75}, 417--420.

\bibitem{Theo:Kout:2008}
\textsc{Theodoridis, S.} \textsc{and} \textsc{Koutroumbas, K.} (2008).
\newblock \emph{Pattern Recognition}, 4 ed.
\newblock Academis Press.

\bibitem{Yao05}
\textsc{Yao, F.}, \textsc{M\"uller, H.~G.}, \textsc{and} \textsc{Wang, J.~L.}
  (2005a).
\newblock Functional data analysis for sparse longitudinal data.
\newblock \emph{J. Am. Statist. Assoc.\/}~\emph{100}, 577--590.

\bibitem{Yao05b}
\textsc{Yao, F.}, \textsc{M\"uller, H.~G.}, \textsc{and} \textsc{Wang, J.~L.}
  (2005b).
\newblock Functional linear regression analysis for longitudinal data.
\newblock \emph{Annals of Statistics\/}~\emph{33}, 2873--2903.

\bibitem{YaoWuZou:16}
\textsc{Yao, F.}, \textsc{Wu, Y.}, \textsc{and} \textsc{Zou, J.} (2016).
\newblock Probability-enhanced effective dimension reduction for classifying
  sparse functional data.
\newblock \emph{Test\/}~\emph{25(1)}, 1--22.

\end{thebibliography}

\end{document}